\documentclass[12pt,showpacs,prd]{revtex4}
\usepackage{epsfig,amssymb,amsfonts}

\makeatletter
\renewcommand{\@makefntext}[1]{\parindent=1em\noindent\hbox to 1.8em{\hss$^{\@thefnmark}$}#1}
\renewcommand{\@footnotemark}{\hbox{\mathsurround=0pt$^{\@thefnmark}$}}
\newcommand{\ftnote}[2]{\footnotemark[#1]\footnotetext[#1]{#2}}
\makeatother
\DeclareMathSymbol{\varGamma}{\mathord}{letters}{"00}
\begin{document}

\title{\hfill{\small\rm FZJ--IKP(TH)--2007--14} \\[1.8em]
Reconciling the $X(3872)$ with the near-threshold enhancement in the
$D^0 \bar D^{*0}$ final state.}

\author{C. Hanhart}
\affiliation{Institut f\"{u}r Kernphysik, Forschungszentrum J\"{u}lich GmbH,
D--52425 J\"{u}lich, Germany}

\author{Yu. S. Kalashnikova, A. E. Kudryavtsev, and A. V. Nefediev}
\affiliation{Institute of Theoretical and Experimental Physics, 117218, B.Cheremushkinskaya 25, Moscow, Russia}

\newcommand{\be}{\begin{equation}}
\newcommand{\bea}{\begin{eqnarray}}
\newcommand{\ee}{\end{equation}}
\newcommand{\eea}{\end{eqnarray}}
\newcommand{\ds}{\displaystyle}

\begin{abstract}
We investigate the enhancement in the $D^0 \bar D^0\pi^0$ final state with
the mass $M=3875.2 \pm 0.7^{+0.3}_{-1.6} \pm 0.8$ MeV
found recently by the Belle Collaboration in the $B \to K D^0 \bar D^0 \pi^0$
decay and test the possibility that
this is yet another manifestation of the well--established resonance $X(3872)$. We
perform a combined Flatt{\`e} analysis of the data for the $D^0 \bar D^0 \pi^0$ mode, and for the $\pi^+ \pi^- J/\psi$ mode of
the $X(3872)$. Only if the $X(3872)$ is a virtual state in the $D^0 \bar D^{*0}$ channel,
the data on the new enhancement comply with those on the $X(3872)$.
In our fits, the mass distribution in the $D^0 \bar D^{*0}$ mode
exhibits a peak at $2\div 3$ MeV above the $D^0 \bar D^{*0}$ threshold, with a distinctive
non-Breit--Wigner shape.
\end{abstract}


\maketitle

\section{Introduction}

The $X(3872$) state, discovered by Belle \cite{Belle} in the $B$-meson
decay, remains the most prominent member of the
family of ``homeless" charmonia, that is those mesons which definitely
contain a $\bar c c$ pair but do not fit the standard
charmonium assignment. The state was confirmed then by CDF
\cite{CDF}, D0 \cite{D0}, and BaBar \cite{BaBar}.
The charmonium option for the $X(3872)$ looks implausible
as the state lies too high to be a $1D$ charmonium, and too
low to be a $2P$ one \cite{BG}. This could, in
principle, mean that we simply do not understand the spectra
of higher charmonia. Indeed, most of the quark model predictions
consider charmonia as $c \bar c$ states in the quark potential model,
with the potential parameters found from the description of lower
charmonia, with uncertainties coming from proper treatment of
relativistic effects. Another source of uncertainty is the role of
open charm thresholds, the problem which is far from being resolved,
though the attempts in this direction can be found in the literature --- see, for example, Refs.~\cite{Eichten,YuSK}.
In any case, it looks premature to reject
the $c \bar c$ assignment for the $X(3872)$ on basis of the mass only.
However, the further development has revealed
more surprises.

The discovery mode of the $X(3872)$ is $\pi^+\pi^- J/\psi$.
The observation of the $X(3872)$ in the
$\gamma J/\psi$ and $\pi^+\pi^-\pi^0 J/\psi$ ($\omega
J/\psi$) modes~\cite{Belle2} implies that the $X$ has positive $C$-parity,
and the dipion in the $\pi^+\pi^- J/\psi$ mode is $C$-odd,
that is it originates from the $\rho$. Coexistence of the $\rho
J/\psi$ and $\omega J/\psi$ modes points to a considerable isospin
violation. Studies of the dipion mass spectrum in
$X(3872) \to \pi^+ \pi^- J/\psi$ decay establish that only
the  $1^{++}$ or $2^{-+}$ quantum number assignments are
compatible with the data, while all other hypotheses are
excluded by more than $3\sigma$ \cite{CDF2}.

Both $1^{++}$ or $2^{-+}$ quantum numbers options for the $X(3872)$
require drastic revisions of naive quark potential models, and
no alternative explanation of the $2^{-+}$ state in this
mass region was suggested. On the other hand,
it was pointed out in Refs.~\cite{VO,RGG} that the $D \bar D^*$ system with
$1^{++}$ quantum numbers can be bound by pion exchange, forming a
mesonic molecule (see also Ref.~\cite{Tornqvist1})\ftnote{1}{An
obvious shorthand notation is
used here and in what follows: $D \bar D^* \equiv \frac{1}{\sqrt{2}}(D \bar
D^* + \bar D D^*)$.}. As confirmed by
actual calculations \cite{Tornqvist2}, large isospin mixing
due to about $8$ MeV difference between the $D^0 \bar
D^{*0}$ and $D^+ D^{*-}$ thresholds can be generated in
the molecular model in quite a natural way. This model was
supplied, in Ref.~\cite{Swanson}, by quark--exchange kernels
responsible for the transitions $D \bar D^* \to \rho J/\psi,
\omega J/\psi$, predicting the $\omega J/\psi$ decay mode of
the $X(3872)$. Note, however, that one-pion-exchange as a binding
mechanism in the $D \bar D^*$ system
should be taken with caution, as, in contrast to $NN$ case, here the pion
can be on-shell, as pointed in \cite{Suzuki}, where the ability to
provide strong enough binding with one-pion exchange was questioned.
For the most recent work on the implications of the nearby pion
threshold see Refs.~\cite{newbraaten,pions}. For recent work for the
$X$ as quark state we refer to Ref.~\cite{meng} and references
therein.

The molecular model has received additional
support with the new data on the mass of the $D^0$
meson \cite{D0mass} which yield a very weak binding,
\be
M_X-M(D^0 \bar D^{*0})= -0.6 \pm 0.6\;{\rm MeV}.
\label{3872mass}
\ee
In the meantime, the Belle Collaboration has reported the
first observation \cite{Bellenew} of the near--threshold
enhancement in the $D^0 \bar D^0 \pi^0$ mode in the decay
$B \to K D^0 \bar D^0 \pi^0$, with the branching fraction
\be
Br(B \to  K D^0 \bar D^0 \pi^0)=(1.22 \pm 0.31^{+0.23}_{-0.30}) \cdot
10^{-4}.
\label{Bellerate}
\ee
The peak mass of the enhancement is measured to be
\be
M_{peak}=3875.2 \pm 0.7^{+0.3}_{-1.6} \pm 0.8\;{\rm MeV}.
\label{Bellemass}
\ee

Obviously it is tempting to relate this new state to the
$X(3872)$. However, the average value of the $X(3872)$ mass is \cite{PDG}
\be
M_X=3871.2 \pm 0.5\;{\rm MeV}.
\ee
The central value (\ref{Bellemass}) of the $ D^0 \bar D^0 \pi^0$ peak mass
enhancement is about $4$ MeV higher than that,
which obviously challenges attempts to relate this new state to the $X(3872)$.

Quite recently, the indication appeared that the Belle result
\cite{Bellenew} is likely to be confirmed.
Namely, the BaBar Collaboration has reported the preliminary data
\cite{BaBarnew} on the
$B \to K D^0 \bar D^{*0}$ decay, where the enhancement with the mass of
\be
M=3875.6 \pm 0.7^{+1.4}_{-1.5} \pm 0.8\;{\rm MeV},
\label{BaBarmass}
\ee
was found, in a very good agreement with (\ref{Bellemass}). BaBar
observes the enhancement in the $D^0 \bar D^0 \pi^0$ {\it and} in the
$D^0 \bar D^0 \gamma$ modes, which strongly supports the presence of
the $D^0 \bar D^{*0}$ intermediate state in the decay of the new $X$.

If the new BaBar data persist, and the enhancement at $3875$ MeV is indeed
seen in two independent experiments, the possibility should be
considered seriously of the presence of two charmonium-like states,
$X(3872)$ and
$X(3875)$, surprisingly close to each other and to the $D^0 \bar D^{*0}$
threshold.

However, there exists another, less exotic possibility. Namely, if the
$X(3872)$ is indeed strongly coupled to the $D^0 \bar D^{*0}$
channel, and indeed has $1^{++}$ quantum numbers, one could
expect the existence of a near--threshold
peak in the $D^0 \bar D^{*0}$ mass distribution.
In the present paper we perform a phenomenological Flatt{\`e}-like
analysis of the data on the decay $B \to K D^0 \bar D^0 \pi^0$ in the near--threshold region under the
assumption of the $X \to D^0 \bar D^{*0} \to D^0 \bar D^0
\pi^0$ decay chain and $1^{++}$ quantum numbers
for the $X$. The data on the $\pi^+ \pi^- J/\psi$ decay modes of
the $X(3872)$ are analyzed in the same framework, in order to investigate
whether these data can accommodate the $X(3875)$ state as a manifestation
of the $X(3872)$.

\section{Flatt{\`e} parametrization}

In this Section we introduce the Flatt{\`e}-like
parametrization of the near--threshold observables.
The relevant mass range is between the thresholds
for the neutral and charged $D$-mesons. A natural
generalization of the standard Flatt{\`e}
parametrization for the near--threshold
resonance \cite{Flatte} of the $D^0 \bar D^{*0}$
scattering amplitude reads
\be
F(E)=-\frac{1}{2k_1}\frac{g_1k_1}{D(E)},
\label{flatte2}
\ee
with
\be
D(E)=\left\{
\begin{array}{ll}
\ds
E-E_f-\frac{g_1\kappa_1}{2}-\frac{g_2\kappa_2}{2}+i\frac{\varGamma(E)}{2},&E<0\\[3mm]
\ds
E-E_f-\frac{g_2\kappa_2}{2}+i\left(\frac{g_1k_1}{2}+\frac{\varGamma(E)}{2}\right),&0<E<\delta\\[3mm]
\ds
E-E_f+i\left(\frac{g_1k_1}{2}+\frac{g_2k_2}{2}+\frac{\varGamma(E)}{2}\right),&E>\delta
\end{array}
\right.
\label{D}
\ee
and
$$
\delta=M(D^+ D^{*-})-M(D^0 \bar D^{*0})=7.6\;{\rm MeV},
$$
$$
k_1=\sqrt{2\mu_1 E},\quad\kappa_1=\sqrt{-2\mu_1 E},\quad k_2=\sqrt{2\mu_2(E-\delta)},\quad\kappa_2=\sqrt{2\mu_2(\delta-E)}.
$$
Here $\mu_1$ and $\mu_2$ are the reduced masses in the $D^0
\bar D^{*0}$ and $D^+ D^{*-}$ channels, respectively,
and the energy $E$ is defined relative to the $D^0 \bar
D^{*0}$ threshold.  In what follows we assume
isospin conservation for the coupling constants, $g_1=g_2=g$.

The term $i\varGamma/2$ in Eq.~(\ref{D}) accounts for
non-$D \bar D^*$ modes. The $X(3872)$ was observed in
the $\pi^+\pi^- J/\psi$, $\pi^+\pi^-\pi^0 J/\psi$, and $\gamma J/\psi$
modes, with
\be
\frac{Br(X \to \pi^+\pi^-\pi^0J/\psi)}{Br(X \to
\pi^+\pi^-J/\psi)}=1.0 \pm 0.4 \pm 0.3,
\label{omega}
\ee
\be
\frac{Br(X \to \gamma J/\psi)}{Br(X \to
\pi^+\pi^-J/\psi)}=0.14 \pm 0.05,
\label{gamma}
\ee
reported in Ref.~\cite{Belle2}. Thus we assume that $\varGamma(E)$ in Eq.~(\ref{D}) is
saturated by the $\pi^+\pi^-J/\psi$ and $\pi^+\pi^-\pi^0J/\psi$
modes and, in accordance with findings of Ref.~\cite{Belle2},
the dipion in the $\pi^+\pi^-J/\psi$ mode comes from
the $\rho$ whereas the tripion in the $\pi^+\pi^-\pi^0J/\psi$ mode
comes from the $\omega$. The $\gamma J/\psi$ channel is neglected due
to its small branching fraction (\ref{gamma}).

The nominal thresholds for both $\rho J/\psi$ and $\omega J/\psi$
($3872$ MeV and $3879$ MeV, respectively) are close to
the mass range under consideration,
but both the $\omega$ meson and, especially, the $\rho$ meson have finite
widths, which are large in the scale under consideration.  Thus $\varGamma(E)$ is
calculated as \be \varGamma(E)=\varGamma_{\pi^+\pi^-J/\psi}(E)+
\varGamma_{\pi^+\pi^-\pi^0J/\psi}(E), \ee \be
\varGamma_{\pi^+\pi^-J/\psi}(E)=f_{\rho}\int^{M-m_{J/\psi}}_{2m_{\pi}}
\frac{dm}{2\pi}\frac{q(m)\Gamma_{\rho}}{(m-m_{\rho})^2+\Gamma_{\rho}^2/4},
\label{rhowidth}
\ee
\be
\varGamma_{\pi^+\pi^-\pi^0J/\psi}(E)=f_{\omega}\int^{M-m_{J/\psi}}_{3m_{\pi}}
\frac{dm}{2\pi}\frac{q(m)\Gamma_{\omega}}{(m-m_{\omega})^2+\Gamma_{\omega}^2/4},
\label{omegawidth}
\ee
with $f_{\rho}$ and $f_{\omega}$ being effective couplings and
\be
q(m)=\sqrt{\frac{(M^2-(m+m_{J/\psi})^2)(M^2-(m-m_{J/\psi})^2)}{4M^2}}
\ee
being the c.m. dipion/tripion momentum ($M=E+M(D^0 \bar D^{*0})$).

Now we are in a position to write down the differential rates in the Flatt{\`e} approximation.
These are
\be
\frac{dBr(B \to K D^0 \bar D^{*0})}{dE}={\cal B}\frac{1}{2\pi}\frac{gk_1}{|D(E)|^2},
\label{DD}
\ee
\be
\frac{dBr(B \to K \pi^+\pi^- J/\psi)}{dE}={\cal B}\frac{1}{2\pi}\frac{\varGamma_{\pi^+\pi^-J/\psi}(E)}{|D(E)|^2},
\label{rhopsi}
\ee
and
\be
\frac{dBr(B \to K \pi^+\pi^-\pi^0 J/\psi)}{dE}={\cal B}\frac{1}{2\pi}\frac{\varGamma_{\pi^+\pi^-\pi^0J/\psi}(E)}{|D(E)|^2}.
\label{omegapsi}
\ee
We assume the short--ranged dynamics of the weak $B \to K$
transition to be
absorbed into the coefficient ${\cal B}$.
Obviously, the rate (\ref{DD}) is defined for $E>0$ only, while
the rates (\ref{rhopsi}) and (\ref{omegapsi}) are defined
both above and below the $D^0 \bar D^{*0}$ threshold.

The formulae (\ref{DD})--(\ref{omegapsi}) are valid in the zero-width approximation for the
$D^*$-mesons. In principle, one could include the finite width of the $D^*$-mesons
either analogous to Eqs.~(\ref{rhowidth}) and
(\ref{omegawidth}) or in a more sophisticated way, as there are
interference effects possible in the final state, as described in Ref.~\cite{Voloshin}.
However, the widths of the $D^*$ mesons are small.
Indeed, the total width of the $D^{*\pm}$-meson is measured to be $96 \pm 22$ keV \cite{PDG}.
There are no data on the $D^{*0}$ width, but one can estimate the $D^0 \pi^0$ width of the $D^{*0}$
from the data \cite{PDG} on charged $D^{*\pm}$, which gives $\Gamma(D^{*0} \to D^0 \pi^0)
= 42$ keV.
The branching fractions of $D^{*0}$
are known \cite{PDG}:
\be
Br(D^{*0} \to D^0 \pi^0)=(61.9 \pm 2.9)\%,
\label{062}
\ee
\be
Br(D^{*0} \to D^0 \gamma) =(38.1 \pm 2.9)\%,
\label{038}
\ee so the total $D^{*0}$ width can be estimated to be only about $68$ keV.
The effect of such a small width was checked to be negligible in our studies,
and we assume the $D^0 \bar D^0 \pi^0$ differential rate to be
\be
\frac{dBr(B\to K D^0 \bar D^0 \pi^0)}{dE}=0.62{\cal B}
\frac{1}{2\pi}\frac{gk_1}{|D(E)|^2},
\label{DDpi}
\ee
where the branching fraction (\ref{062}) is taken into account.

Analogously we have for the $D^0 \bar D^0 \gamma$ differential
rate
\be
\frac{dBr(B\to K D^0 \bar D^0 \gamma)}{dE}=0.38{\cal B}
\frac{1}{2\pi}\frac{gk_1}{|D(E)|^2}.
\label{DDgamma}
\ee
Expressions (\ref{DDpi}) and (\ref{DDgamma}) neglect final--state
interactions; in particular, no $D
\bar D$ resonance within a few MeV above $D^0 \bar D^0$ threshold is
assumed to exist, and $\pi$-rescattering is neglected. The latter is
expected to be weak, as a consequence of chiral symmetry \cite{pions}.

\section{Flatt{\`e} analysis: procedure and results}

Let us first specify the data used in our analysis.
For the $\pi^+ \pi^- J/\psi$ mode we use the data from the $B$-meson decay.
These are the ones reported by the Belle \cite{Belle} and
BaBar \cite{BaBardata} Collaborations.
The $X(3872)$ is seen by Belle in the charged $B$-meson decay, with $35.7
\pm 6.8$ signal events, and with the branching fraction \cite{Belle}
\be
Br(B^+ \to K^+ X)Br(X \to \pi^+ \pi^- J/\psi)=(13.0 \pm 2.9 \pm 0.7)
\cdot 10^{-6}.
\label{rhochargedbelle}
\ee
The BaBar Collaboration \cite{BaBardata} has observed the $X(3872)$
both in the charged and
neutral $B$-meson decays, with $61.2 \pm 15.3$ signal events
for the
charged mode, and only $8.3 \pm 4.5$ signal events for the neutral one.
The branching fraction for the charged mode was found to
be
\be
Br(B^- \to K^- X)Br(X \to \pi^+ \pi^- J/\psi)=(10.1 \pm 2.5 \pm 1.0) \cdot
10^{-6},
\label{rhochargedbabar}
\ee
while the result for the neutral mode is much less certain: a $90$\%
confidence interval was established as
\be
1.34 \cdot 10^{-6}<Br(B^0 \to K^0 X)Br(X \to \pi^+ \pi^- J/\psi)<10.3 \cdot
10^{-6}.
\label{rhoneutral}
\ee
Due to large errors and much smaller number of events, the $X(3872)$ peak
in the neutral mode looks much less convincing than the peak in the
charged mode.

A similar situation takes place for the  $D^0 \bar D^0 \pi^0$ final state.
The Belle data \cite{Bellenew} include both $B^+ \to K^+ D^0 \bar D^0
\pi^0$ and  $B^0 \to K^0 D^0 \bar D^0 \pi^0$ decays. There are $17.4 \pm
5.2$ signal events in the charged mode, with the branching fraction
\be
Br(B^+ \to K^+ D^0 \bar D^0 \pi^0)=(1.02 \pm 0.31^{+0.21}_{-0.29}) \cdot
10^{-4},
\label{dcharged}
\ee
and $6.5 \pm 2.6$ signal events in the neutral mode, with
\be
Br(B^0 \to K^0 D^0 \bar D^0 \pi^0)=(1.66 \pm 0.70^{+0.32}_{-0.37}) \cdot
10^{-4}.
\label{dneutral}
\ee
Data on the $B^+$ and $B^0$ decays separately are presented in
Ref.~\cite{Belletalk}. The $D^0 \bar D^0 \pi^0$ enhancement appears to be
clearly seen in the data on charged $B$ decays while, again, the neutral
mode displays, within the errors, a much less pronounced peak.

We conclude therefore that the data on charged and neutral $B$ decays
should be analyzed separately. The present analysis is performed for the charged mode only.
Namely, with the Flatt{\`e} formalism, we attempt to describe simultaneously
the $\pi^+ \pi^- J/\psi$ mass spectrum
from the charged mode and the $D^0 \bar D^0 \pi^0$ spectrum
from the $B^+$ mode, taken from Ref.~\cite{Belletalk}.

The branching fractions (\ref{rhochargedbelle}) and
(\ref{rhochargedbabar}) differ but, within the errors, are consistent
with each other. In both sets of data, the fitted width of the signal
is consistent with the resolution, so only the upper limits on the
$X(3872)$ width were established:
\be
\Gamma_{tot}({\rm Belle}) < 2.3 \;{\rm MeV}
\label{bellewidth}
\ee
and
\be
\Gamma_{tot}({\rm BaBar}) < 4.1 \;{\rm MeV},
\label{babarwidth}
\ee
for the Belle and BaBar data, respectively. In view of this discrepancy we
prefer to present two sets of fits, based on the two aforementioned sets of the $\pi^+
\pi^- J/\psi$ data.

The $\pi^+ \pi^- J/\psi$ data are fitted in the interval $-20<E<20$ MeV
(as before, $E$ is the energy relative to the $D^0 \bar D^{*0}$ threshold),
after subtraction of the full background found in the corresponding analysis. The free
parameters of the fit are the short--range factor $\cal B$ and the Flatt{\`e}
parameters $E_f$, $g$, and $f_{\rho}$. The parameter $f_{\omega}$ is
constrained, in accordance with Eq.~(\ref{omega}) through the condition
\be
\frac{R_{\rho J/\psi}}{R_{\omega J/\psi}}=1,
\ee
where
\be
R_{\rho J/\psi}=\int^{20 MeV}_{-20 MeV}\frac{dBr(B \to K
\pi^+\pi^- J/\psi)}{dE}dE,
\label{Rrho}
\ee
\be
R_{\omega J/\psi}=\int^{20 MeV}_{-20 MeV}\frac{dBr(B \to K
\pi^+\pi^-\pi^0 J/\psi)}{dE}dE.
\label{Romega}
\ee
The limits of integration in Eqs.~(\ref{Rrho}) and (\ref{Romega}) are
somehow arbitrary but, as most of the support of the distributions
(\ref{rhopsi}) and (\ref{omegapsi}) comes from within a few MeV around
the $D^0 \bar D^{*0}$ threshold, the uncertainty introduced by the limits
of integration is much less than the experimental errors in
Eq.~(\ref{omega}).

The $D^0 \bar D^0 \pi^0$ data are fitted in the energy region $0<E<20$ MeV.
Equation (\ref{DDpi}) describes the production of the $D^0 \bar D^0
\pi^0$ mode via the $X$-resonance, while the $D\bar D^*$ pairs are known to be
copiously produced in the $B \to K$ decay in a non-resonant way. Besides, the
$D^0 \bar D^0 \pi^0$ final state could come from non-$D^0 \bar D^{*0}$ modes
like, for example, $B \to K^* D^0 \bar D^0$. Therefore,
we are to make assumptions on the background.

The background in Refs.~\cite{Bellenew} and \cite{Belletalk} is mostly
combinatorial, and this part, given explicitly in the publications, was subtracted
prior to the analysis.
For the rest of the background it is not possible to separate
the contributions of the $D^0 \bar D^{*0}$ and the $D^0 \bar D^0 \pi^0$ due
to a limited phase space~\cite{Bellenew}. So we work under two extreme
assumptions for the background. In Case A we consider the $D^0 \bar
D^0\pi^0$ background as unrelated to the $D^0 \bar D^{*0}$ channel,
while in Case B we assume that all the $D^0 \bar D^0\pi^0$ events come
from the $D^0 \bar D^{*0}$ mode.
The background was evaluated by fitting the Belle data off--peak
($25<E<50$ MeV).
In Case A the background function is assumed to be proportional to the
three--body $D^0
\bar D^0\pi^0$ phase space $R_3\propto E_{DD\pi}^2$, where
 $E_{DD\pi}=E+m_{D^{*0}}-m_{D^0}-m_{\pi^0}$.
Then the total $B \to K D^0 \bar D^0\pi^0$ differential rate is
\be
\frac{dBr^A(B \to K D^0 \bar D^{0}\pi^0)}{dE}=0.62\frac{{\cal B}}{2\pi}\frac{gk_1}{|D(E)|^2}+c_AE_{DD\pi}^2,
\label{tot3}
\ee
with $c_A$ as fitting constant. In Case B the background function
is proportional to the two--body $D^0 \bar D^{*0}$ phase space
$R_2\propto k_1$ (see the definition below Eq.~(\ref{D})). Then the
signal--background interference is to be taken into account:
\be
\frac{dBr^B(B \to
K D^0 \bar D^{0}\pi^0)}{dE}= 0.62\frac{k_1}{2\pi}\left[\left({\rm Re}\frac{\sqrt{g \cal B}}{D(E)}
+c_B\cos\phi\right)^2 +\left({\rm Im}\frac{\sqrt{g \cal B}}{D(E)}+c_B\sin\phi\right)^2\right],
\label{tot2}
\ee
with the relative phase $\phi$ and $c_B$ being fitting constants.

The differential rates are translated into number-of-events distributions
as follows. There are about $36$ signal events in the Belle data, which
corresponds to the branching fraction of about $1.3 \cdot 10^{-5}$ (see Eq.~(\ref{rhochargedbelle})).
Thus the number-of-events per $5$ MeV distribution for the
$\pi^+ \pi^- J/\psi$ mode is given by
\be
N^{\pi \pi J/\psi}_{\rm Belle}(E)=5 \,  {\rm [MeV]} \left(\frac{36}{1.3 \cdot
10^{-5}}\right)
\frac{dBr(B \to K \pi^+ \pi^- J/\psi)}{dE} .
\label{rhonoebelle}
\ee
For the BaBar data, with $61$ events and the branching fraction of about $1.02 \cdot 10^{-5}$ (see Eq.~(\ref{rhochargedbabar})), we have
\be
N^{\pi \pi J/\psi}_{\rm BaBar}(E)=5 \,  {\rm [MeV]} \left(\frac{61}{1.02 \cdot
10^{-5}}\right)
\frac{dBr(B \to K \pi^+ \pi^- J/\psi)}{dE} .
\label{rhonoebabar}
\ee
As to the $D^0 \bar D^0 \pi^0$ mode, the Belle Collaboration states to
have $17.4$ signal events in the charged mode
\cite{Bellenew}, which corresponds to the
branching fraction (\ref{dcharged}) of about
$1.02 \cdot 10^{-4}$.  The
number-of-events distributions per $4.25$ MeV
for the $D^0 \bar D^0 \pi^0$ mode
is calculated as
\be
N_{A,B}^{D^0 \bar D^0 \pi^0}(E)=4.25 {\rm [MeV]} \left(\frac{17.4}{1.02 \cdot 10^{-4}}\right)
\frac{dBr^{A,B}(B \to K D^0 \bar D^{0}\pi^0)}{dE} .
\label{dnoe}
\ee

The best fit to the  $\pi^+ \pi^- J/\psi$ data alone requires
a vanishing value of the $D \bar D^*$ coupling constant, $g=0$, so
that such solution cannot accommodate the $D^0 \bar D^0 \pi^0$
enhancement as a
related phenomenon. To describe both  $\pi^+ \pi^- J/\psi$ and $D^0 \bar
D^{*0}$ modes we are to compromise on the  $\pi^+ \pi^- J/\psi$ line--shape.

\begin{table}[t]
\caption{The set of the Flatt{\`e} parameters for the best fits to the
Belle data Ref.~\cite{Belle} and \cite{Bellenew}.}
\begin{ruledtabular}
\begin{tabular}{ccccccc}
Fit&$g$&$f_\rho$&$f_\omega$&$E_f$, MeV&${\cal B}
\cdot 10^4$&$\phi$\\
\hline
\hline
$A_{\rm Belle}$&0.3&0.0070&0.036&-11.0&11.0&---\\
\hline
$B_{\rm Belle}$&0.3&0.0086&0.046&-10.9&8.9&-144$^0$\\
\end{tabular}
\end{ruledtabular}
\label{tfitsbelle}
\end{table}
\begin{table}[t]
\caption{The set of the Flatt{\`e} parameters for the best fits to the
BaBar data of Ref.~\cite{BaBardata} and the Belle data of Ref.~\cite{Bellenew}.}
\begin{ruledtabular}
\begin{tabular}{ccccccc}
Fit&$g$&$f_\rho$&$f_\omega$&$E_f$, MeV&${\cal B}
\cdot 10^4$&$\phi$\\
\hline
\hline
$A_{\rm BaBar}$&0.3&0.0042&0.021&-8.8&11.4&---\\
\hline
$B_{\rm BaBar}$&0.3&0.0056&0.027&-8.8&8.9&-153$^0$\\
\end{tabular}
\end{ruledtabular}
\label{tfitsbabar}
\end{table}

It appears that a decent combined fit can be achieved only if the $\pi^+ \pi^-
J/\psi$ distribution is peaked {\it exactly} at the $D^0 \bar
D^{*0}$ threshold, with the peak width (defined as the width at the
peak half--height) close to the upper limits
given by  Eq.~(\ref{bellewidth}) or (\ref{babarwidth}).
The values of the coupling $g$ were found to be of the order of magnitude or larger than $0.3$.
Finally, the fits exhibit the scaling behaviour: they
remain stable under the transformation
\be
g \to \lambda g,\quad E_f \to \lambda E_f,\quad f_{\rho} \to \lambda f_{\rho},
\quad f_{\omega} \to \lambda f_{\omega},\quad {\cal B} \to \lambda {\cal B},
\label{scaling}
\ee
with tiny variations of the phase $\phi$ in the Case B.

\begin{figure}[t]
\begin{tabular}{|c|c|c|c|}
\hline
\epsfig{file=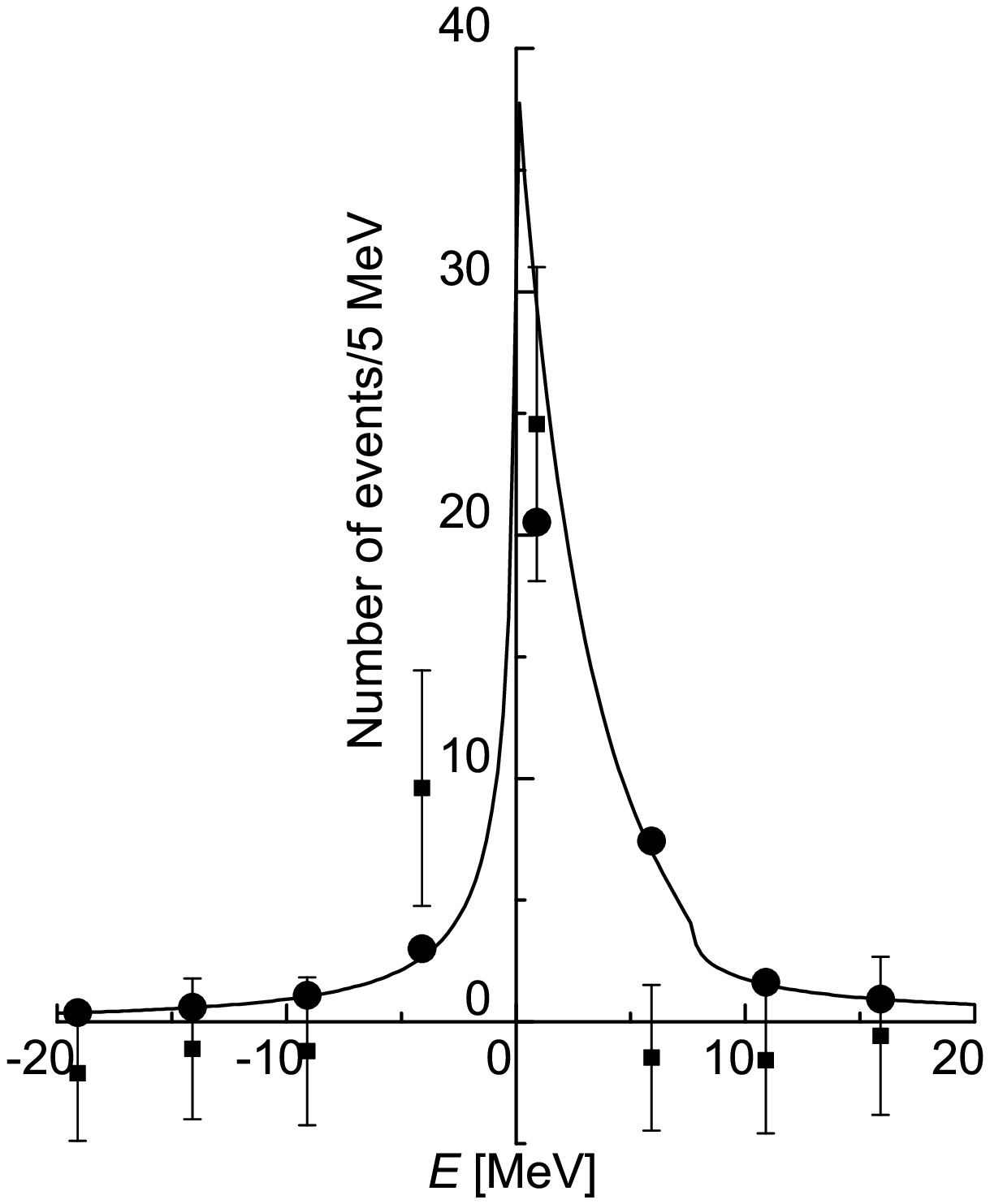,width=3.9cm}&\epsfig{file=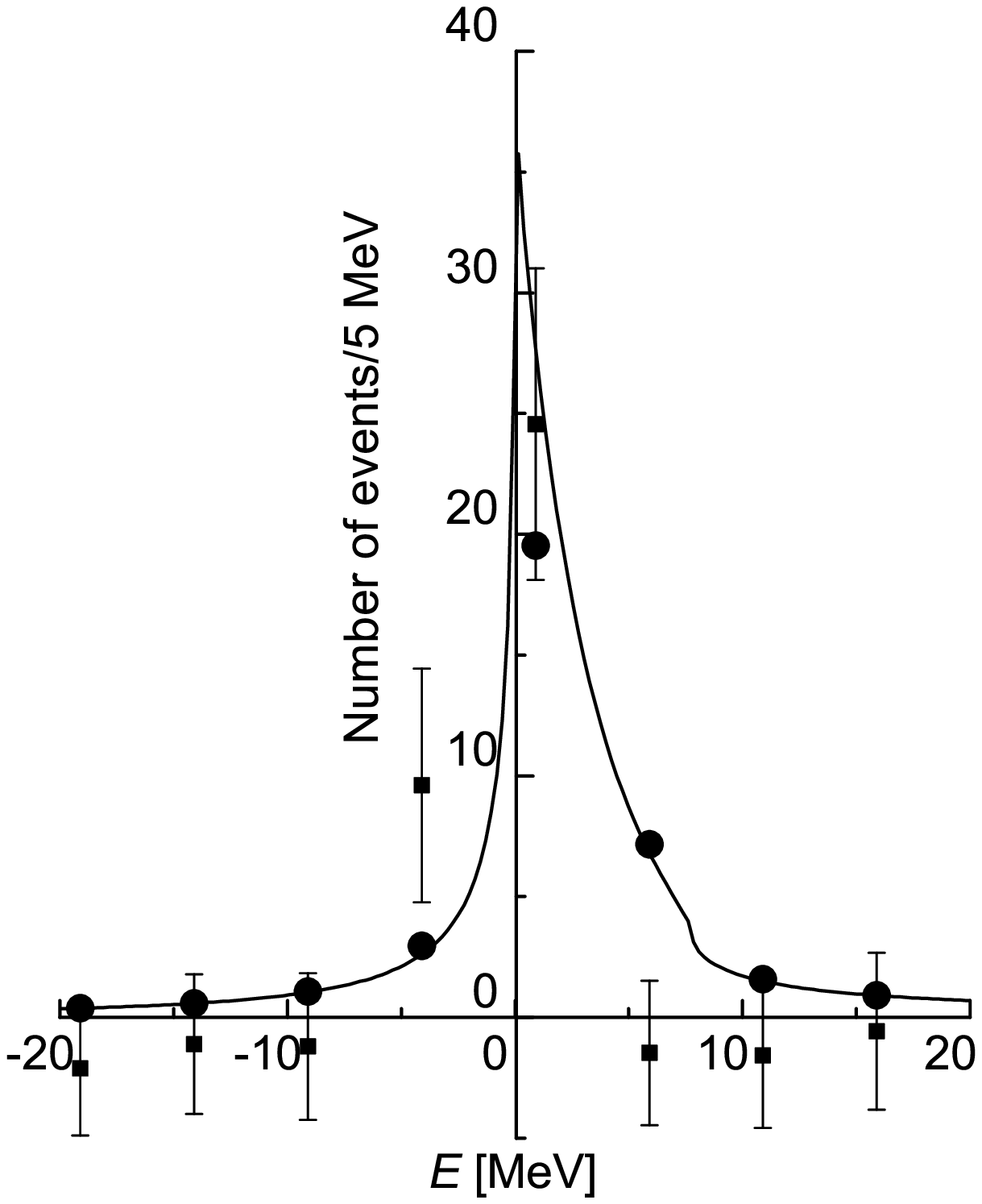,width=3.9cm}&
\epsfig{file=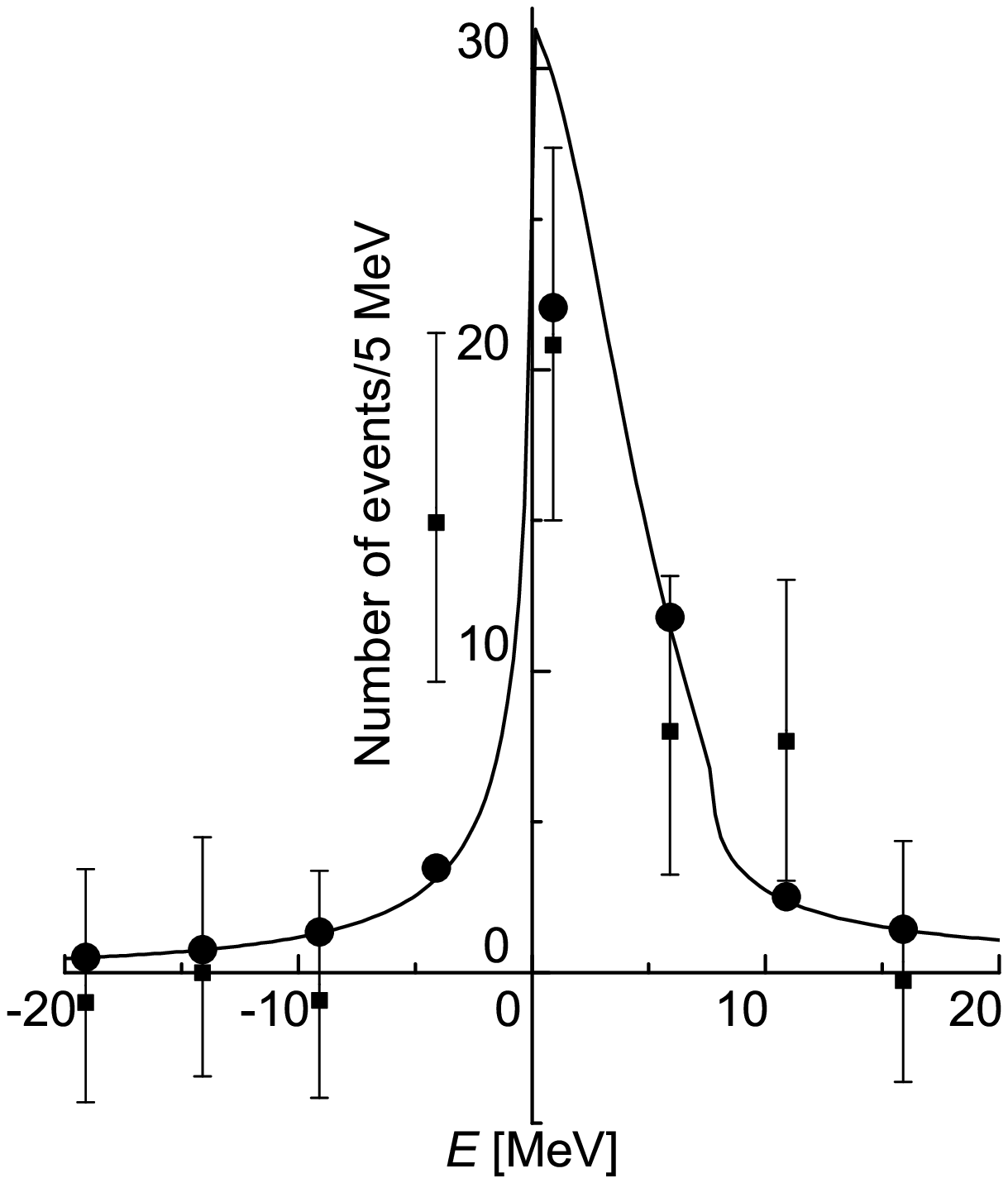,width=3.9cm}&\epsfig{file=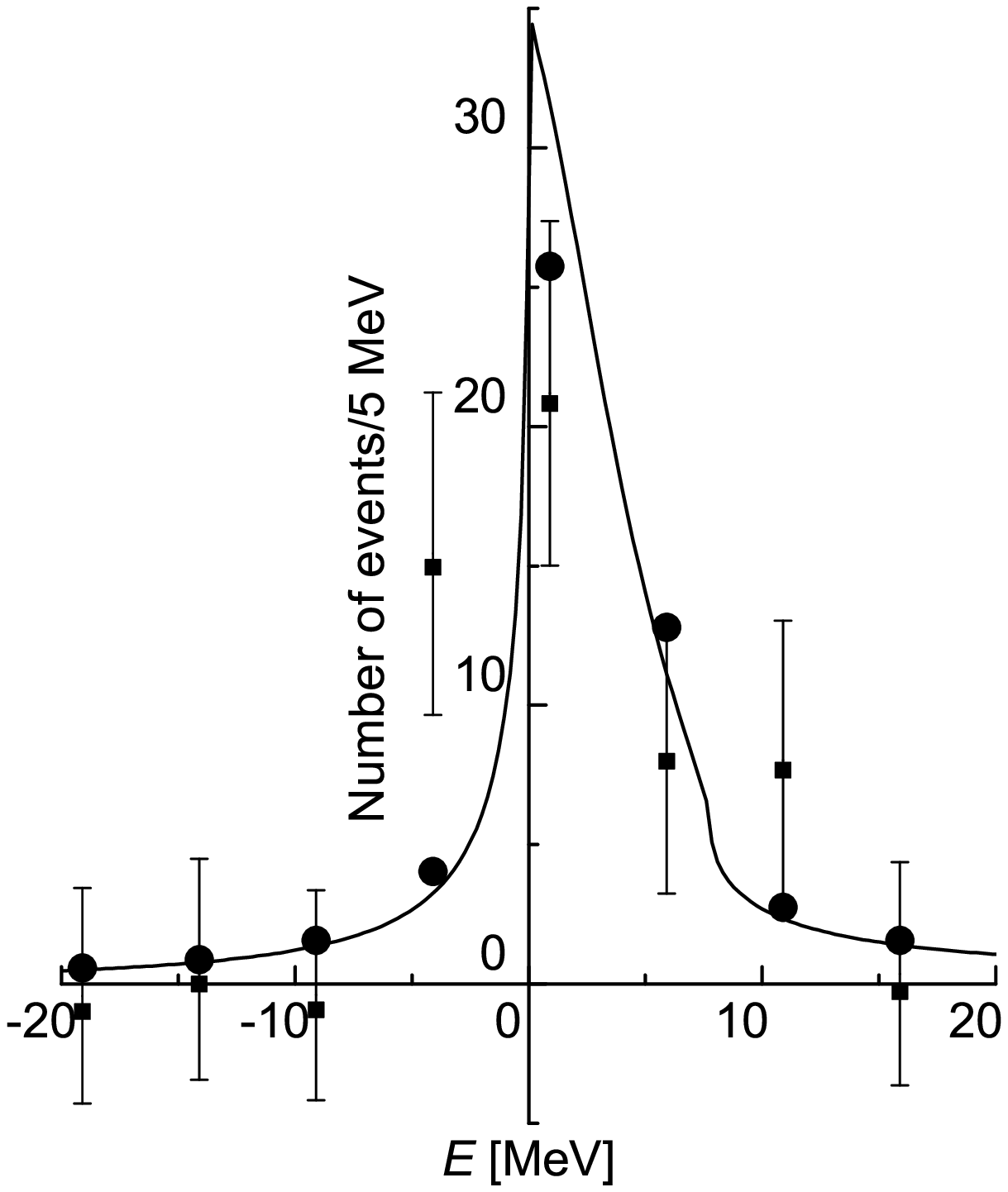,width=3.9cm}\\
\hline
\epsfig{file=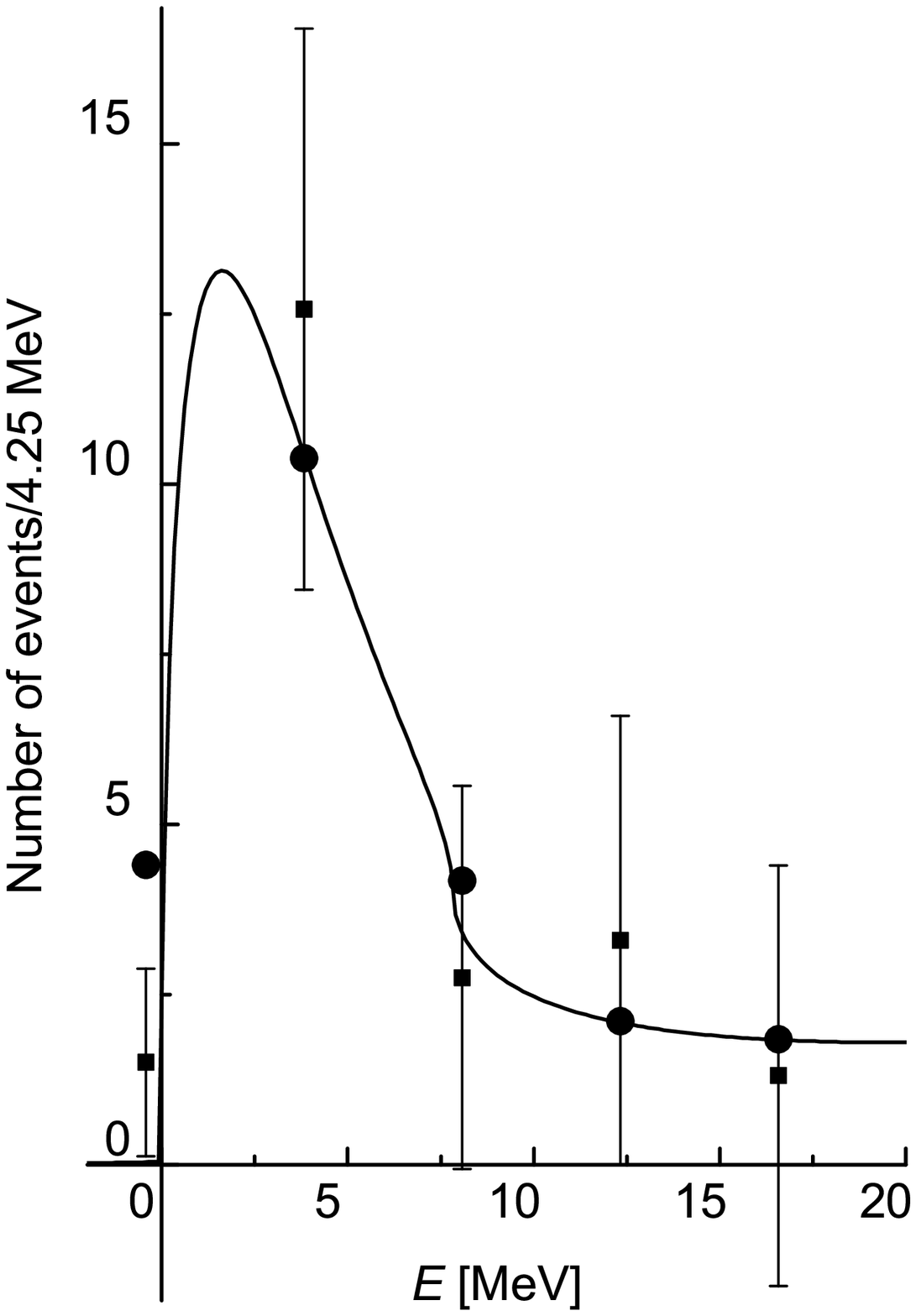,width=3.9cm}&\epsfig{file=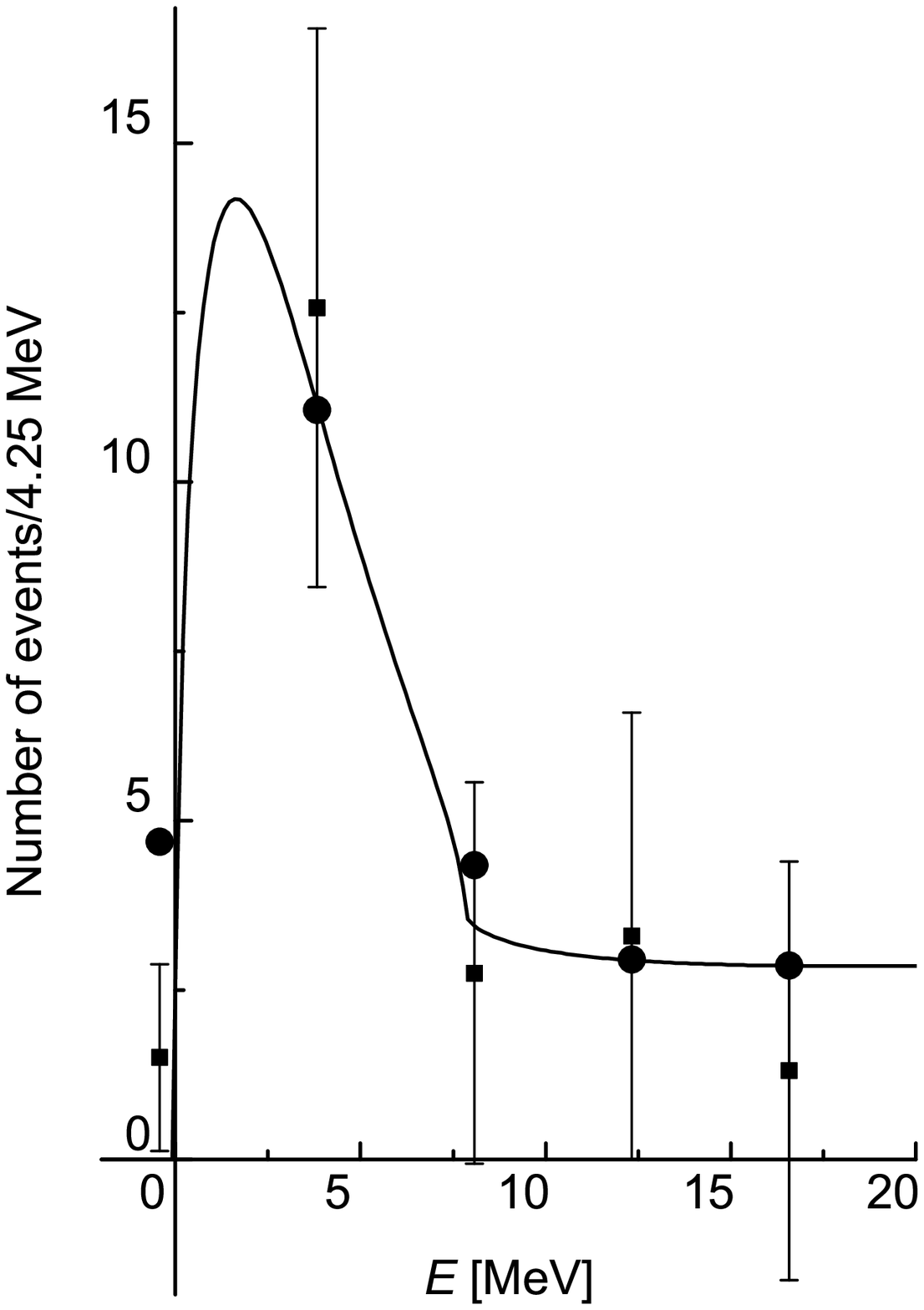,width=3.9cm}&
\epsfig{file=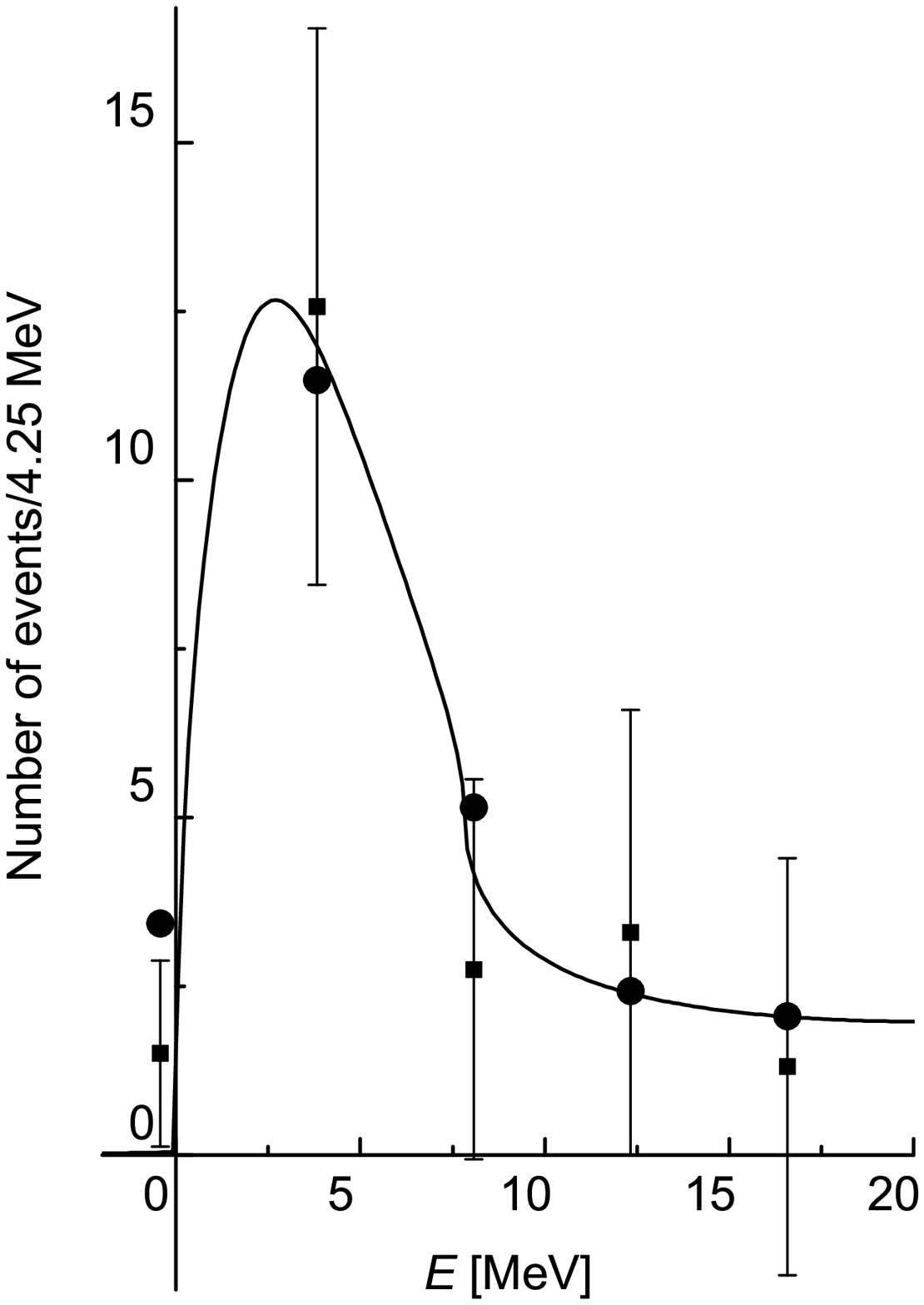,width=3.9cm}&\epsfig{file=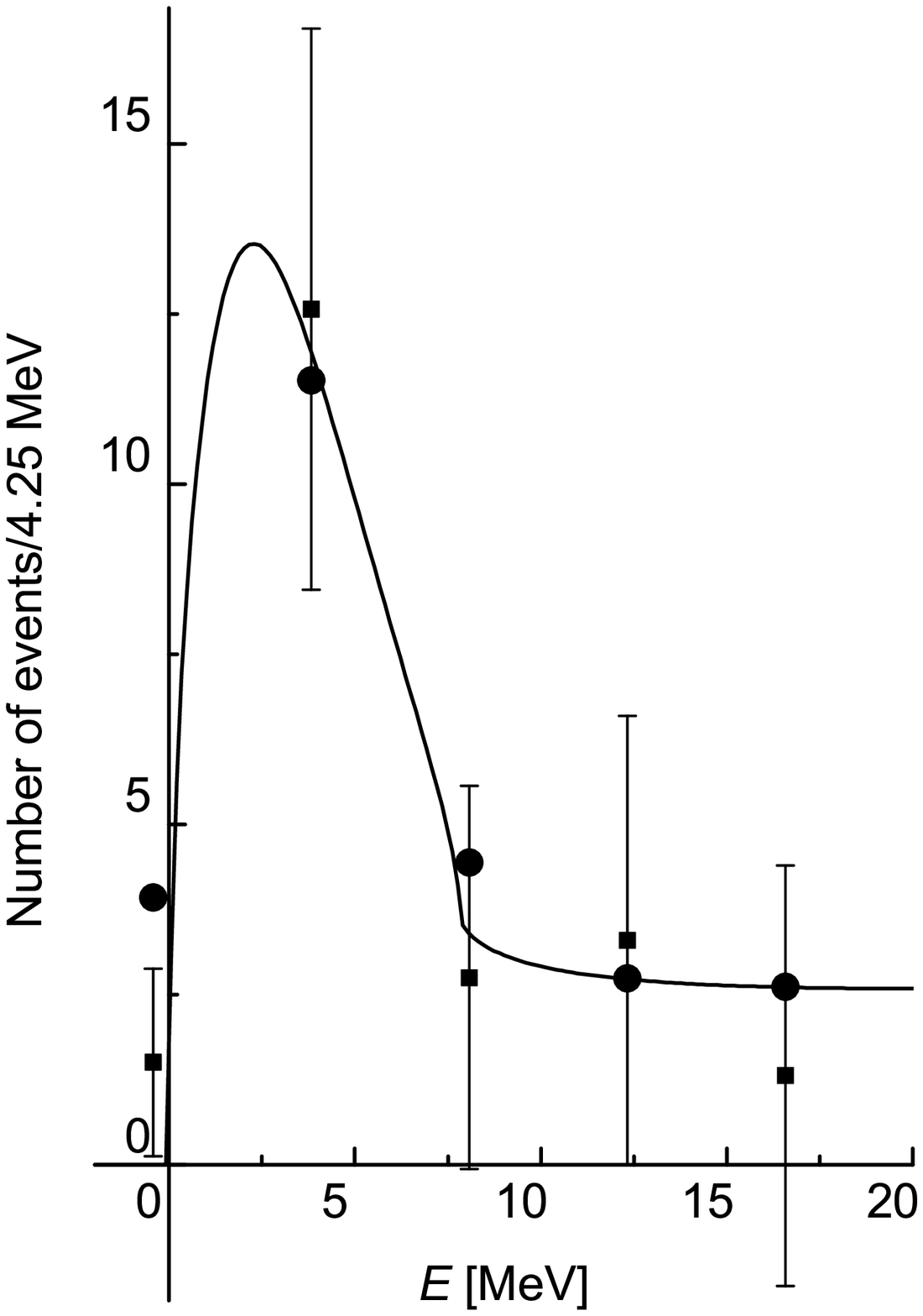,width=3.9cm}\\
\hline
$A_{\rm Belle}$&$B_{\rm Belle}$&$A_{\rm BaBar}$&$B_{\rm BaBar}$\\
\hline
\end{tabular}
\caption{Upper plots: Our fits to the differential rates for the $\pi^+ \pi^- J/\psi$
channel measured by Belle~\cite{Belle} and BaBar~\cite{BaBardata}
using prescription A and B (see Eqs.~(\ref{tot3}) and (\ref{tot2})).
Lower plots: Corresponding fits for the differential rates in the $D^0 \bar D^0 \pi^0$ channel measured by Belle \cite{Belletalk}.
The distributions integrated over the bins are shown in each panel as filled dots, experimental data as filled squares with
error bars.}
\label{fig1}
\end{figure}

In Tables~\ref{tfitsbelle},~\ref{tfitsbabar} we present the
sets of the best fitting parameters --- for both Case A and Case B and for
$g=0.3$ --- for the Belle (Table~\ref{tfitsbelle}) and BaBar (Table~\ref{tfitsbabar}) data
on the $\pi^+ \pi^- J/\psi$ mode and for the Belle data for the $D^0 \bar D^0 \pi^0$ mode.
To assess the quality of the fits we calculate the
$\pi^+ \pi^- J/\psi$ distributions integrated over the $5$ MeV bins,
as in Refs.~\cite{Belle} and \cite{BaBardata}, and the $D^0 \bar D^0 \pi^0$
distributions integrated over the $4.25$ MeV bins, as in Refs.~\cite{Bellenew} and
\cite{Belletalk}. The results are shown at Fig.~\ref{fig1} together with the
experimental data.

The above--mentioned scaling behaviour does not allow one to perform a
proper fit with the estimate of uncertainties in the parameters found.
Indeed, the parameters of the best fits found for the values of coupling constant $g$
larger than $0.3$ differ only by a few \% from the ones given by the scaling
transformation (\ref{scaling}), and the corresponding distributions are
very similar to those given at Fig.~\ref{fig1}.

As seen from the figures, acceptable fits require the $D^0 \bar D^{*0}$
differential rate to be peaked at
around $2\div 3$ MeV above the $D^0 \bar D^{*0}$ threshold.
The scattering length in the $D^0 \bar D^{*0}$ channel  which follows
from the expression (\ref{flatte2}) of the $D^0 \bar D^{*0}$ scattering
amplitude, is given by the expression
\be
a=-\frac{\sqrt{2\mu_2\delta}+2E_f/g+i\varGamma(0)/g}{(\sqrt{2\mu_2\delta}+2E_f/g)^2+\varGamma(0)^2/g^2},
\ee
and is calculated to be
\be
a=\left\{
\begin{array}{ll}
(-3.98-i0.46)~{\rm fm},&{\rm Case~A_{\rm Belle}}\\
(-3.95-i0.55)~{\rm fm},&{\rm Case~B_{Belle}},
\end{array}
\right.
\label{scatbelle}
\ee
and
\be
a=\left\{
\begin{array}{ll}
(-3.10-i0.16)~{\rm fm},&{\rm Case~A_{\rm BaBar}}\\
(-3.10-i0.22)~{\rm fm},&{\rm Case~B_{BaBar}}.
\end{array}
\right.
\label{scatbabar}
\ee
The real part of the scattering length for all the fits appears to be
large and negative, and the imaginary part is much smaller. This,
together with the beautiful cusp in the $\pi^+ \pi^- J/\psi$
mass distribution, signals the presence of a virtual state
in the $D^0 \bar D^{*0}$ channel. The cusp scenario for the
$\pi^+ \pi^- J/\psi$ excitation curve in the $X(3872)$ mass range
was advocated in Ref.~\cite{Bugg}. The $X(3872)$ as a virtual $D \bar D^*$ state
was found in the coupled--channel microscopic quark model \cite {YuSK}.

A large scattering length explains naturally the scaling behaviour of
the Flatt{\`e} parameters. Such kind of scaling was
described in Ref.~\cite{AEKflatte} in the context of light scalar
mesons properties: the scaling behaviour occurs if the
scattering length approximation is operative.
In the case of $X$ the situation is more complicated, as there are two
near--threshold channels, neutral and charged. Nevertheless,
if it is possible to neglect the energy
$E$ in the expression (\ref{D}) for the Flatt{\`e} denominator
$D(E)$ then, as seen from the expression (\ref{flatte2}),
scaling for the $D^0 \bar D^{*0}$ scattering amplitude
indeed takes place. If the factor $\cal B$ obeys the scaling
transformation, the differential rates
(\ref{DD})--(\ref{omegapsi})
also exhibit the scaling behaviour. Note that, if the energy
dependence of the charged $D^+ D^{*-}$ and non-$D \bar D^*$ channel contributions
is neglected as well, this corresponds to the scattering length
approximation, and neglect of the effective radius term.

\section{Discussion}

\begin{figure}[t]
\centerline{\epsfig{file=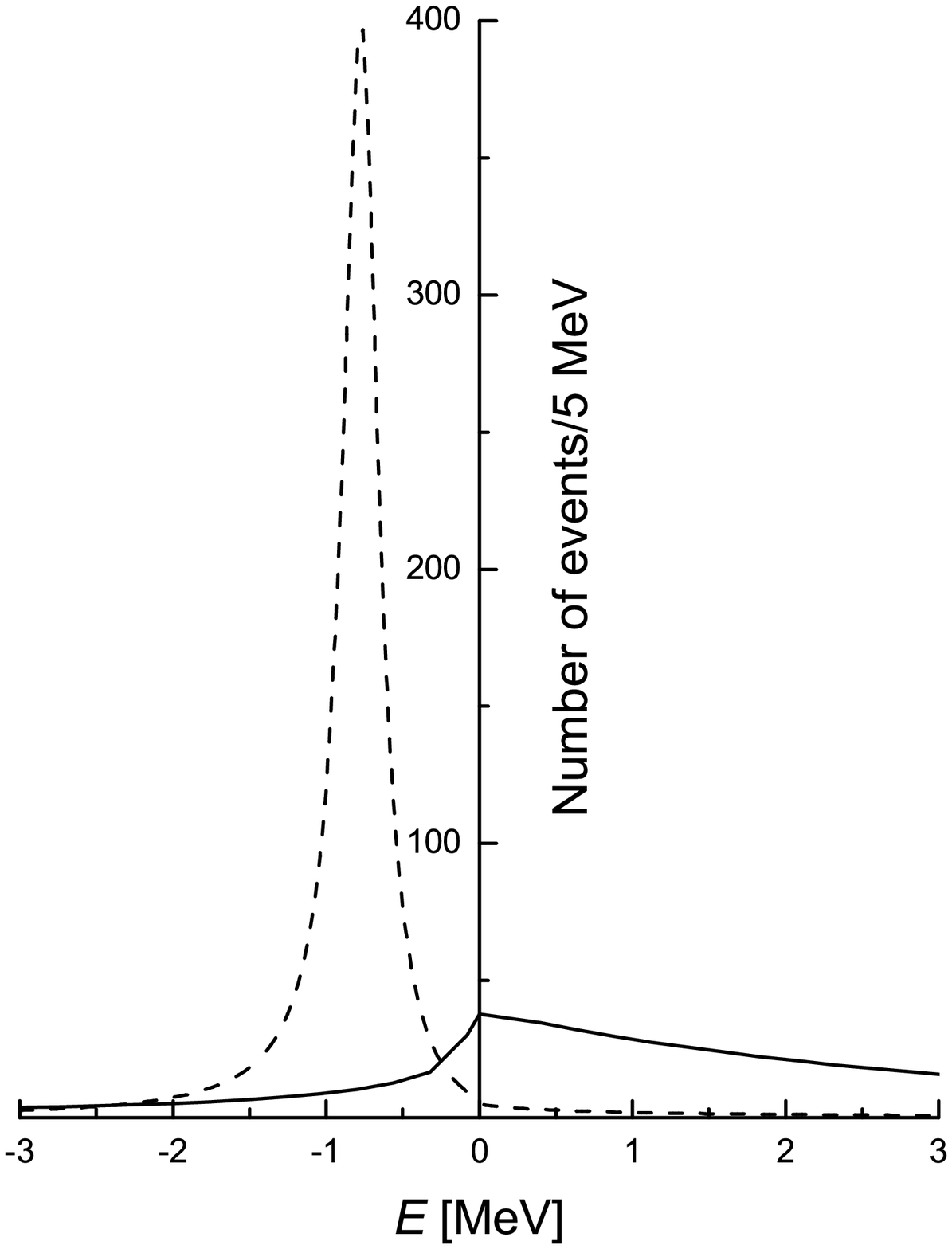,width=6cm}\hspace*{1cm}
\epsfig{file=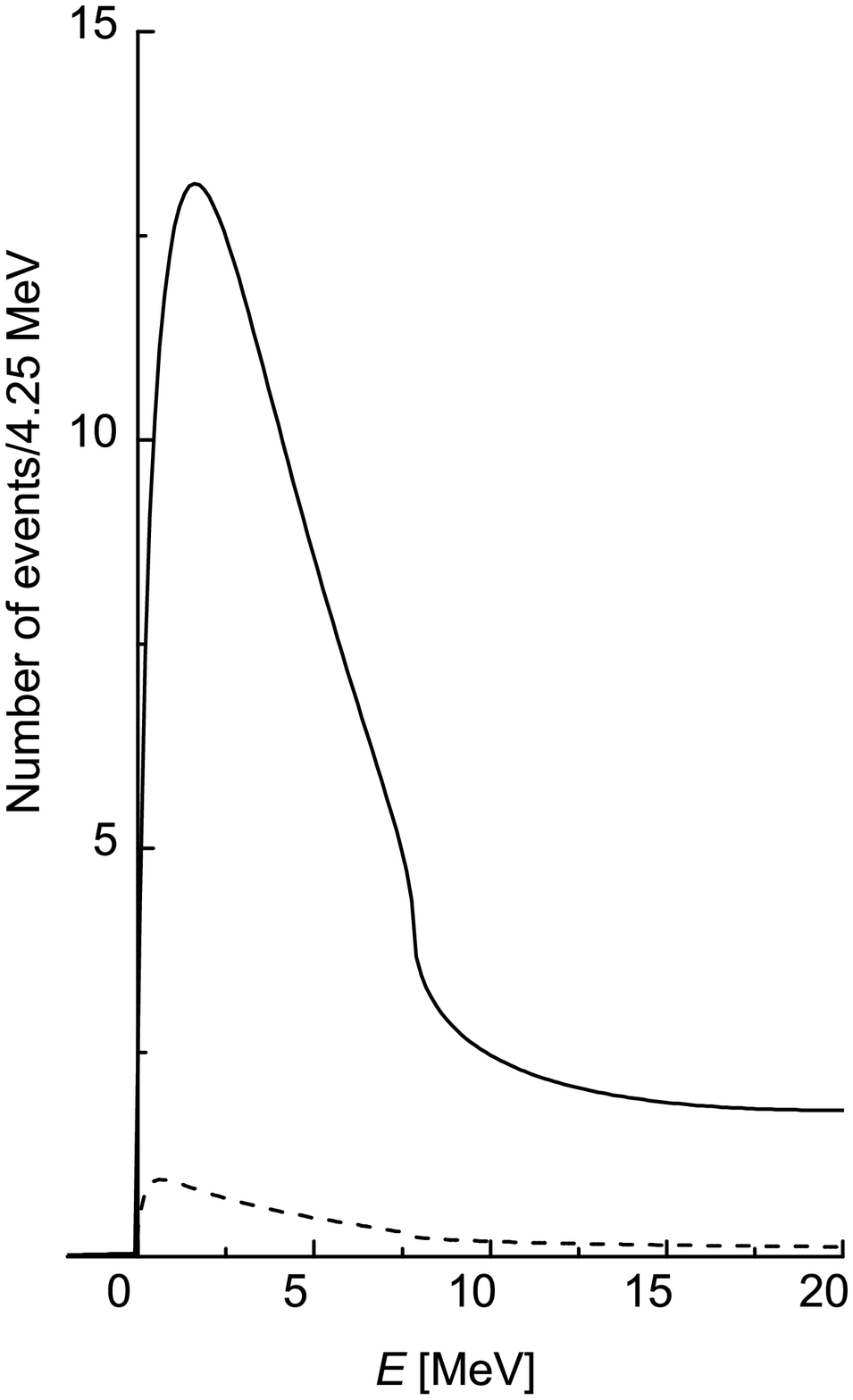,width=5cm}}
\caption{The differential rates for the $\pi^+ \pi^- J/\psi$ (first
plot) and $D^0
\bar D^{*0}$ (second plot) for the fits $A_{\rm Belle}$ (solid curves) and
$C$ (dashed curves).}
\label{figbound}
\end{figure}

Our analysis shows that the large branching fraction (\ref{Bellerate})
implies the $X$ to be a virtual $D^0 \bar D^{*0}$ state,
and not a bound state.  We illustrate this point by calculating the rates
(\ref{DD}) and (\ref{rhopsi}) for the set of the Flatt{\`e} parameters (fit
$C$)
\be
g=0.3,\quad E_f=-25.9\;{\rm MeV},\quad f_{\rho}=0.007,\quad
f_{\omega}=0.036, \quad {\cal B}=1.32 \cdot 10^{-4}.
\label{bound}
\ee
The values of the coupling constants coincide with those of the fit
$A_{\rm Belle}$,
while the parameter $E_f$ is chosen to yield the real part of the scattering
length to be equal in magnitude to the one evaluated for the given fit
$A_{\rm Belle}$, but positive: $\tilde{a}=(+3.98-i0.46)$ fm. The parameter
$\cal B$ for this set yields the same value of the total branching
fraction
for the $\pi^+ \pi^- J/\psi$ mode as the fit $A_{\rm Belle}$.
The $\pi^+ \pi^- J/\psi$ and $D^0 \bar D^0 \pi^0$ rates are shown in Fig.~\ref{figbound},
together with the
rates obtained for the case $A_{\rm Belle}$ (without background).  The new
curve (dashed line in Fig.~\ref{figbound}) displays a very narrow peak in the $\pi^+ \pi^- J/\psi$ distribution,
corresponding to the $D^0
\bar D^{*0}$ bound state, with binding energy of about $1$ MeV (there is no
corresponding peak in the $D^0 \bar D^0 \pi^0$ distribution as the finite
width of the $D^{*0}$ is not taken into account in our analysis). Note that
the $\pi^+ \pi^- J/\psi$ rates (Fig.~\ref{figbound}) are normalized to
give the branching
ratio $1.3 \cdot 10^{-5}$, which requires the coefficient ${\cal B}$ to be
much larger for the virtual state than for the bound state. As a result, the
$D^0 \bar D^{*0}$ rate is much smaller for the bound state, as seen from
Fig.~\ref{figbound}.

Obviously, the difference between the bound--state and
virtual-state cases for the ratio
\be
\frac{Br(X \to D^0 \bar D^0 \pi^0)}{Br(X \to \pi^+\pi^- J/\psi)}
\label{r}
\ee
is driven by the strength of the bound--state peak,
as discussed in Ref.~\cite{Braaten}, where the scattering length
approximation was used to describe the $X(3872)$.
Following Ref.~\cite{Braaten}, let us write down the scattering length in the
$D^0 \bar D^{*0}$ channel as
\be
a=\frac{1}{\gamma_{re}+i\gamma_{im}}.
\ee
Then, in the scattering length approximation, the
$\pi^+\pi^-J/\psi$ differential rate is proportional to the factor
\begin{eqnarray}
\ds\frac{\gamma_{im}}{\gamma_{re}^2+(k_1+\gamma_{im})^2},&E>0,\nonumber\\
\label{braaten}\\[-3mm]
\frac{\gamma_{im}}{(\gamma_{re}-\kappa_1)^2+\gamma_{im}^2},&E<0,\nonumber
\end{eqnarray}
while the $D^0 \bar D^{*0}$ rate is proportional to
\be
\frac{k_1}{\gamma_{re}^2+(k_1+\gamma_{im})^2}.
\ee
The line--shape for the $D^0 \bar D^{*0}$
channel does not depend on the sign of $\gamma_{re}$. The same is true for the
$\pi^+\pi^-J/\psi$ line--shape above the $D^0 \bar D^{*0}$ threshold while, below the threshold,
the line--shapes differ drastically: in the bound--state case there is a narrow peak
below threshold, and in the virtual-state case a threshold cusp appears.

For $\gamma_{re}>0$ and $\gamma_{im} \to 0$ the expression (\ref{braaten})
becomes a $\delta$-function (see Ref.~\cite{Braaten}):
\be
\frac{\pi}{\mu_1}\gamma_{re}\delta(E+\gamma_{re}^2/(2\mu_1)) .
\ee
Then the total rate does not depend on $\gamma_{im}$, if it is small enough.  This
simply means that, for $\gamma_{im}=0$, we have a real bound state, which is
not coupled to inelastic channels.  In contrast to the bound--state case, for
the virtual state, the rate (\ref{braaten}) tends to zero with $\gamma_{im} \to 0$,
while the $D^0 \bar D^{*0}$ rate does not vanish in such a limit. So it is
possible, adjusting $\gamma_{im}$, to obtain large values of the ratio
(\ref{r}).

Exactly the same situation is encountered in our fit: we need $g
\gtrsim 0.3$ for the
fit to be reasonable and, in this scaling regime, as soon as we have a
positive real part of the scattering length, the ratio (\ref{r}) becomes
small while, with a negative real part, we get a solution compatible with the
data. The large branching fraction (\ref{Bellerate}) was identified in
Ref.~\cite{Swansonreport} as a disaster for the molecular model of the $X(3872)$.
Indeed, the bound--state molecule decay into
$D^0 \bar D^0 \pi^0$ is driven by the
process $D^{*0} \to D^0 \pi^0$ which gives the width of order $2\Gamma(D^{*0} \to D^0 \pi^0)$
(up to the interference
effects calculated in Ref.~\cite{Voloshin} which, for the bound--state case,
cannot be neglected anymore and should be taken into account). The main
decay mode of the $X$ is $\pi^+ \pi^- J/\psi$ because the phase
space available is large. This is confirmed by model
calculations of Ref.~\cite{Swanson} yielding
\be
\frac{Br(X \to D^0 \bar D^0 \pi^0)}{Br(X \to \pi^+\pi^- J/\psi)} \approx 0.08,
\label{Sdpsi}
\ee
in a strong contradiction with data.

The estimate (\ref{Sdpsi}) describes the decay of an isolated bound
state. However, the suppression is more moderate as, in $B$-decay, the
continuum contribution is also to be considered. The bound--state
contribution would  be zero in the zero--width approximation for $D^{*0}$, while
the $D^0 \bar D^{*0}$ continuum contribution remains finite if the $D^{*0}$ width is
neglected.  However, if the $X$ is a bound state, the continuum contribution
is not large (see Fig.~\ref{figbound}),
\be
\frac{Br(X \to D^0 \bar D^0 \pi^0)}{Br(X \to \pi^+\pi^- J/\psi)}
\approx 0.62.
\label{dpsibound}
\ee
Such a small rate would remain unnoticed against the
background. So, in practice, the bound--state
$X(3872)$ would reveal itself only as a narrow peak below
threshold, with a very small rate (see Eq.~(\ref{Sdpsi})).
In contrast to this we get for the virtual state
\be
\frac{Br(X \to D^0 \bar D^0 \pi^0)}{Br(X \to \pi^+\pi^- J/\psi)}
\approx 9.9.
\label{dpsivirt}
\ee

In our analysis,
the $X$ appears to be a virtual state in the $D^0 \bar D^{*0}$ channel.
This
does not contradict the assumption $g_1=g_2=g$ employed in the analysis.
The latter means that the underlying strong interaction
conserves isospin, and all the isospin violation comes from the mass
difference between charged and neutral $D \bar D^*$ thresholds. No
charged
partners of the $X$ are observed, so it is reasonable to assume that
the strong attractive interaction takes place in the isosinglet $D
\bar D^*$
channel.

We do not specify the nature of this attractive force.
It is known that in the one-pion-exchange model for the $X$, the force is
attractive in the isosinglet channel, and is repulsive in the isotriplet
one. However, as was already mentioned, the doubts were cast in
\cite{Suzuki} on the role of one-pion-exchange in the $D \bar D^*$
binding, and it was advocated there that the $X$ may fit the $2^3P_1$
charmonium assignment if the coupling to $D \bar D^*$ channel is taken
into
account. In such a scenario the strong binding force obviously takes
place in the isosinglet channel.

We note, however, that, with the Flatt{\`e} parameters found,
one can make a definite statement: whatever the
nature of the $X(3872)$ is, the
admixture of a compact $c \bar c$ state in its wavefunction is small.
Both large scattering length and the scaling behaviour of the $D \bar
D^{*}$
amplitude are consequences of the large value of the coupling
constant of the state to the $D \bar D^*$
channel. As shown in Ref.~\cite{evidence}, this
points to a large $D \bar D^*$ component and a
dynamical origin of the $X$. Although
formulated for quasi--bound states in Ref.~\cite{evidence} the
argument can also be generalized to virtual states.
To clarify the connection between effective coupling
and the nature of the state
observe that the two--point
function $g(s)$ for the resonance can be written as
\be
g(s)=\frac{1}{s-M^2-i\bar \Sigma(s)},
\label{resonance}
\ee
where $M$ is the physical mass of the resonance and $\bar \Sigma(s) =\Sigma(s)-$Re$\Sigma(M^2)$
is the self--energy responsible for
the dressing through the mesonic channels. In the near--threshold
region the momenta involved are much smaller than the inverse of
the range of forces. As a result one may neglect
the $s$--dependence of the real part of $\bar \Sigma$
and replace its imaginary part by the leading terms
\be
g(s)\simeq \frac{1}{s-M^2+iM\sum_ig_ik_i},
\label{nearthreshold}
\ee
where the sum is over near--threshold channels, and the contributions of
distant thresholds are absorbed into the
renormalised mass $M$. Nonrelativistic
reduction of Eq.~(\ref{nearthreshold}) immediately yields the Flatt{\`e}
formula (\ref{flatte2}). Thus the Flatt{\`e} parameter $E_f$
acquires clear physical meaning: the quantity $M(D^0 \bar D^{*0}) + E_f$
is the physical mass of the resonance, renormalised by the coupling to the
decay channels.

Now, if the couplings $g_i$ are small, the distribution for the resonance
takes a standard Breit--Wigner form, and the scattering length is small.
Correspondingly, the state is mostly $c \bar c$, with a small admixture of the
$D \bar D^*$ component.
If the couplings are large, the terms proportional to $g_ik_i$ control
the denominator in Eq.~(\ref{nearthreshold}), the Breit--Wigner shape is
severely distorted, the scattering length
approximation is operative, and the mesonic component dominates
the near--threshold wavefunction.

Formulated differently: if the couplings are large, the properties
of the resonance are given mainly by the continuum contribution ---
which is equivalent to saying it is mostly of molecular (dynamical) nature.
It should be stressed that this kind of reasoning can only be used,
if the resonance mass is very close to a threshold, for then the
contribution of the continuum state is dominated by the unitarity
cut piece which is unique and model independent.
This argument is put into more quantitative terms in Ref.~\cite{evidence}.
It is also important to note that our analysis does not
allow for any conclusion on the mechanism that leads to the
molecular structure.
On the level of the phenomenological
parametrisations used here a molecule formation due to
$t$--channel exchanges and due to short--ranged $s$--channel forces
($c \bar c$--$DD^*$ mixing)
would necessarily lead to the same properties of the state, once
the parameters are adjusted to the data.

\section{Summary}

In this paper we present a Flatt{\`e} analysis of the
Belle data \cite{Bellenew} on the near--threshold enhancement
in the $D^0 \bar D^0 \pi^0$ mode. We constrain the
Flatt{\`e} parametrization with the data on the $X(3872)$
seen in the $\pi^+ \pi^- J/\psi$ and $\pi^+ \pi^- \pi^0 J/\psi$ modes.
With such constraints the new state can be understood as a
manifestation of the well--established $X(3872)$ resonance.

We showed that the structure at $3875$ MeV can only be related to the
$X(3872)$, if we assume the $X$ to be of a dynamical origin, however,
not as a bound state but as a virtual state. The situation is then
similar to that of nucleon--nucleon scattering in the spin--singlet
channel near threshold: in contrast to the spin--triplet channel,
where there exists the deuteron as a bound state,
the huge scattering length in the spin--singlet channel --- about 20 fm ---
comes from a near--threshold virtual state. The attractive
interaction is just not strong enough to form a bound state
in this channel as well.

The line--shape in the $D^0 \bar D^{*0}$ mode appears to differ
substantially from the one extracted previously from the
Belle data directly. It peaks much closer to the $D^0 \bar D^{*0}$
threshold, though the overall description of the data looks
quite reasonable within the experimental errors.

It is the $\pi^+ \pi^- J/\psi$ line--shape which, in our solutions, differ
drastically from the one described by a simple Breit--Wigner form. We
found a threshold cusp, with a width close to the limits imposed by the data
analysis.
While the data currently available allows for such a line--shape, a
considerable improvement in the experimental resolution could confirm or rule
out this possibility. In the meantime, we urge to perform an analysis of the data on the
$D^0 \bar D^0 \pi^0$ final state with Flatt{\`e} formulae given
in Eqs.~(\ref{DD})--(\ref{omegapsi}).

Equally important is the Flatt{\`e} analysis of the $D^0 \bar D^0 \gamma$ data \cite{BaBarnew}: if the structure in the
$D^0 \bar D^0 \pi^0$ is indeed due to $D^0 \bar D^{*0}$ and
is indeed related to the $X(3872)$ as a virtual state, one
should observe an enhancement in $D^0\bar D^0 \gamma$ similar to the one seen in the $D^0 \bar D^0 \pi^0$.
The phase space available in this final state is larger than that in
$D^0 \bar D^0 \pi^0$, so it is easier to separate the contributions
of $D^0 \bar D^{*0}$ and $D^0 \bar D^0 \gamma$ to the peak.
The  $D^0 \bar D^0 \gamma$ enhancement would be
described with the Flatt{\`e} formula (\ref{DDgamma}) and, up to
background and possible FSI effects, the
ratio of branching fractions would be
\be
\frac{Br(X \to D^0 \bar D^0 \pi^0)}{Br(X \to D^0 \bar D^0 \gamma)}\approx 1.6.
\ee

The most interesting situation would happen if, due to an improved
resolution in the  $\pi^+ \pi^- J/\psi$ mode, the
combined Flatt{\`e} analysis of the $\pi^+ \pi^- J/\psi$, $D^0 \bar D^0 \pi^0$, and
$D^0 \bar D^0 \gamma$ data fails to deliver a self-consistent result.
Such a situation would point to the new $X(3875)$ state being completely
unrelated to the $X(3872)$.

\acknowledgments

We would like to thank A. Dolgolenko for useful comments and suggestions and P. Pakhlov for illuminating
discussions on various aspects of the Belle experiment.
This research was supported by the Federal Agency for Atomic
Energy of Russian Federation, by the grants RFFI-05-02-04012-NNIOa, DFG-436 RUS
113/820/0-1(R), NSh-843.2006.2, and NSh-5603.2006.2, and
by the Federal Programme of the Russian Ministry of
Industry, Science, and Technology No. 40.052.1.1.1112. A.N. is also supported through the
project PTDC/FIS/70843/2006-Fisica.

\end{document}